\documentclass[conference]{IEEEtran}
\IEEEoverridecommandlockouts
\usepackage{amsmath,amsfonts}
\usepackage{bbm}
\usepackage{algorithm}
\usepackage{algorithmic}
\usepackage{multirow}
\usepackage{listings}%
\usepackage{tabularx}
\usepackage{textcomp}
\usepackage{stfloats}
\usepackage{url}
\usepackage{verbatim}
\usepackage{amsthm}
\usepackage{graphicx}
\usepackage{bbm}
\usepackage{cite}
\usepackage{color}

\UseRawInputEncoding  
\usepackage{subcaption}
\usepackage{booktabs}
\usepackage{caption}
\usepackage{subcaption}

\renewcommand{\algorithmicrequire}{\textbf{Input:}}

\begin{document}
\title{Efficient Multi-Worker Selection based Distributed Swarm Learning via Analog Aggregation
\thanks{This work was supported in part by the U.S. National Science Foundation grants \#2146497, \#2231209, \#2244219, \#2315596,  \#2343619, \#2413622 and \#2425811; and in part by a grant from BoRSF under the contract LEQSF(2024-27)-RD-B-03.}
}
\author{\IEEEauthorblockN{Zhuoyu Yao$^\dagger$, \; Yue Wang$^\dagger$, \; Songyang Zhang$^*$, \; Yingshu Li$^\dagger$, \; Zhipeng Cai$^\dagger$, \; Zhi Tian$^\star$}
\IEEEauthorblockA{$^\dagger$Department of Computer Science, Georgia State University, Atlanta, GA, USA\\$^*$Department of Electrical and Computer Engineering, University of Louisiana at Lafayette, LA, USA\\$^\star$Department of Electrical and Computer Engineering, George Mason University, Fairfax, VA, USA}}

\maketitle

\begin{abstract}
Recent advances in distributed learning systems have introduced effective solutions for implementing collaborative artificial intelligence techniques in wireless communication networks. 
Federated learning approaches provide a model-aggregation mechanism among edge devices to achieve collaborative training, while ensuring data security, communication efficiency, and sharing computational overheads. 
On the other hand, 
limited transmission resources and complex communication environments  remain significant 
bottlenecks to 
efficient collaborations among 
edge devices, particularly 
within 
large-scale networks. 
To address such issues, this paper proposes an over-the-air (OTA) analog aggregation method 
designed for the 
distributed swarm learning (DSL), termed DSL-OTA, aiming to enhance 
communication efficiency, enable effective cooperation, and ensure privacy preserving. 
Incorporating multi-worker selection strategy with over-the-air aggregation not only makes the standard DSL based on single best worker contributing to global model update to become more federated, but also secures the aggregation from potential risks of data leakage.  
Our theoretical analyses verify the advantages of the proposed DSL-OTA algorithm in terms of fast convergence rate and low communication costs.
Simulation results reveal that our DSL-OTA outperforms the other existing methods by achieving better learning performance under both homogeneous and heterogeneous dataset settings.

\end{abstract}


\section{Introduction}
With the rapid growth of mobile device and wireless communication techniques, mobile edge networks perform as an imperative workhorse 
for Internet of Things (IoTs). Recently, distributed swarm learning (DSL)~\cite{wang2024distributed} raises as a 
biological intelligence (BI) empowered artificial intelligence (AI) solution 
for wireless communication networks. However, standard DSL is originally determined by the best worker selection mechanism~\cite{fan2023cb}, deviating from the theme of collaborative learning such as Federated learning (FL)~\cite{mcmahan2016federated, fan2022ICC}. 
The partial cooperation hinders the comprehensive utilization of distributed, multi-modal and numerous sensing data, 
leading to insufficient understanding of patterns and unstable learning processes, impeding the generalization and robustness of DSL in large-scale wireless networks. 


In the literature, 
probabilistic device selection strategies have been developed to collaboratively train the global model with 
affordable communication overhead~\cite{shahid2021communication}. 
Such method allows 
the central server to draw a comprehensive picture for local performance and distributed sensing records, 
improving the collaboration gain of distributed learning in edge networks~\cite{zhang2022communication}. 
However, 
those collaboration strategies inevitably introduce extra communication overheads,
as the uploading requirements of multiple participants 
requires more communication resources such as increased total transmission power and bandwidth. 
Next, the non-ideal wireless environments and complex physical layer networks post 
an urgent challenge for reliable  design of communication model  over wireless networks. 
Moreover, the communication overhead is exacerbated  
in data heterogeneity scenarios at the edge, 
in which the asynchronous system incurs 
escalating communication costs, model updating latency and the lack of adaptivity by stochastic gradient descent (SGD)-based model updates. The aforementioned technical flaws hinder the effective and consistent information transmission in the deployment of distributed learning solutions on wireless networks.


For the effective communication model design, the analog computation mechanism has gained attentions in 
distributed learning. 
The significantly high multi-access utilization rate and light computational cost validate the communication efficiency in wireless environment~\cite{csahin2023survey, fan2023icc}. 
While 
improved performances have been achieved by enhanced  system design, 
adaptive model optimization and aggregation~\cite{reddi2020adaptive}, 
sparsity-aware compression~\cite{fan20221}, and effective communication scheduling strategy~\cite{zhang2022communication}, 
these algorithm-level designs 
fails in low latency transmission, 
which is more significant in large-scale, complex edge wireless communications. Some communications theory-based designs benefits in mitigating the complexity of communication scheme with source multiplexing prototypes, 
An insightful solution termed 
over-the-air (OTA) aggregation performs as the reliable channel modeling and more efficient channel multiplexing for multi-access wireless communication. OTA introduces an multiple access channel (MAC) to permit the multiplexing of single wireless channel~\cite{yang2020federated}. The waveform superposition properties of MAC 
serves as a superposed channel utilization rule to allow multiple access share the same channel with the same time slot and frequency, benefiting in radio resources allocation, data privacy insurance and low cost communication. Nevertheless, the deployment of OTA under the framework of distributed swarm learning systems still needs further exploration for theoretical proof of communication efficiency and model convergence. 

For efficient communication model design, this paper investigates  the specific wireless communication challenges for massive users scenarios and the effective information transmission scheme in multi-worker selection DSL, by developing new DSL techniques based on multi-worker selection and over-the-air aggregation. 
Our main contributions are summarized as follows.

\begin{enumerate}
\item We propose an efficient distributed swarm learning (DSL) framework with over-the-air (OTA) aggregation for high communication efficiency in wireless edge networks, named DSL-OTA. To the best of our knowledge, this is the first work that integrates DSL with OTA techniques customized for wireless edge networks under the constrains of limited communication resources.

\item Further, we design a joint optimization for multi-worker selection and power allocation under the DSL-OTA framework, which not only improves the performance of standard DSL but also enhances the communication efficiency of edge networks at fast convergence speed in model training. 

 \item 
 Theoretically, we also analyze the convergence behavior and communication costs of our proposed DSL-OTA, which indicates that the proposed DSL-OGA algorithm converges as fast as the existing FL-OTA methods at the cost of less communication costs while achieving better accuracy and converging faster than the standard DSL.    
 


\end{enumerate}
\emph{Notations:} $\|\cdot\|$ represents the Euclidean norm of a vector or a matrix. The expectation and the first order derivative are represented by $\mathbb{E}$ and $\nabla$. $\odot$ is Hadamard product. The probability of an event $y=c$ is expressed as $\Pr(y=c)$. The event indicator is designed as $\mathbbm{1}_{y=c}$, which is equal to $1$ when $y=c$, or $0$ otherwise.

\section{Communication Efficient Distributed Swarm Learning}

This section develops the problem formulation 
DSL-OTA.
The illustration is started with the problem statement of common distributed learning tasks in wireless edge networks. 
Next, we demonstrate the proposed DSL-OTA model from multi-worker selection strategy and OTA communication model based joint optimization mechanism for adaptive collborative learning in the edge. 
Then, the algorithm of DSL-OTA is detailed for the IoT deployment.

\subsection{Problem Formulation} \label{Problem Formulation}
In wireless application of edge computing, the mobile edge AI consists of the learning model and 
communication model to cooperate for 
a general pattern on 
distributed datasets. Consider a cross-device wireless network setup with $C$ workers, each 
contains the local dataset ${D_i} = \left\{ {\left( {{x_{i,k}},{l_{i,k}}} \right)} \right\},{x_{i,k}} \in {\cal X},{l_{i,k}} \in {\cal L}$ and the synthetic global dataset $D_g$ for worker evaluation. 
Within $t$-th communication round, the  
broadcasts the global model $ M\left( {{\mathbf{w}}_{t}^{\overline{g}},D_g} \right)$ for individual training. Local models $ M_i\left( {\mathbf{\mathbf{w}}_{i,t},D_i} \right)$ are trained by $D_i$ and get the assessment of their contribution $F_{i,t\left( {\mathbf{\mathbf{w}}_{i,t},D_i} \right)}$. 
The communication 
model is achieved by the over the air computation, as the allowed workers transmit model parameter vector $\mathbf{w}_{i,t}=\{w_{i,t}^1, \ldots,w_{i,t}^n\} \in {\cal R}^N$ within the same MAC synchronously. The global model $\mathbf{w}_{i,t}^{\overline{g}}$ is updated by the aggregated gain of personal optimal models $\{\mathbf{w}_{i,t}\}$, that are natural summed with the demodulation result $\mathbf{w}_{i,t}$. The optimum solution represents 
an unbiased approximation of the mapping function $f: \mathcal{X} \to \mathcal{L}$ for all the local datasets.

\subsection{Muti-Worker Selection based DSL Model}

For the effective collaboration and frequent information interaction, we incorporate the multi-worker selection strategy to alleviate the expensive overheads of all model uploads. 
DSL~\cite{wang2024distributed, fan2023cb} serves 
as an efficient and robust solution for data heterogeneous IoTs.  
Vanilla DSL define a scalar function based on the Root Mean Square Error (RMSE) for 
local training evaluation, as
\begin{flalign}
    && {F_i}\left( {{\mathbf{w}_{i,t+1}};{D_i}} \right) &= \frac{1}{{\left| {{D_i}} \right|}}\sum\limits_{({x_{i,j}},{l_{i,j}}) \in {D_i}} \!\!\!\!\!{\sqrt {{{\left( {M({\mathbf{w}_{i,t}},{x_{i,j}}) \!- {l_{i,j}}} \right)}^2}} }. & \label{eq:RMSE}  
    \end{flalign}
For the symbols simplicity, $F_{i,t + 1},F_{i,t+1}^g$  is used to represent ${F_i}\left( {{\mathbf{w}_{i,t+1}};{D_i}} \right),{F_i}\left( {{\mathbf{w}_{i,t+1}};{D_g}} \right)$ hereafter.

Compared with standard DSL, multi-worker selection based DSL enforces 
a network intelligence strategy to strengthen the collaborative scheme across diverse participants. 
Different from the best worker updating mechanism, 
multi-worker selection strategy explores a data quality empowered individual evaluation for the consistency assessment of data heterogeneity and learning performance as: 
\begin{flalign}
    &&  \theta_{i,t} &=  {{\tau}F\left( {{\mathbf{w}_{i,t}},{D_i}} \right) + {\left(1 - \tau\right)}\eta _{i}} ,&\label{eq:weighted_constrain}
\end{flalign}
where the hyperparameter ${\tau}$ contributes to the trade-off between the data quality and the learning capacity. $\eta _{i}$ represents the consistency of $i$-th dataset with global dataset. The multi-worker selection strategy propose a more reliable approach to explore the most efficient workers to participate in global learning.

To thrive the patterns of global aggregation, we design the worker selection strategy as the integer programming with the optimization objective for the largest set of workers participating in pattern recognition in distributed learning process, as 
\begin{flalign}
    &&   J\left( {{S_t}} \right) &= \max{\sum\limits_{i = 1}^C {{s_{i,t}}} }   & \label{eq:objective} \\
    &&  \mbox{s.t., }\; \theta_{i,t}s_{i,t} &\le \overline \theta  _{t-1} ,i = 1, \ldots ,C,  \nonumber
\end{flalign}
where $S_t=\{s_{1,t},\ldots, s_{C,t}\},s_i \in \{0,1\}$ is the selected subset of participants, $s_{i,t}=1$ reveals the $i$-th worker is allowed to upload its parameters in $t$-th round. $\textstyle{{{\bar \theta }_{t - 1}} = \frac{1}{C}\sum\limits_{i \in C} {{\theta _{i,t - 1}}}}$ is the adaptive threshold to ensure the selected workers perform better than the average scale of last round. 

Thus, the local parameter updates of individual training and the global parameter updates are modified as 
\begin{flalign}
    &&  {\mathbf{w}_{i,t + 1}} \!=& \mathbf{w}_{i,t} + {c_0}{v_{i,t}} + {c_1}\left( {\mathbf{w}_{i.t}^l - {\mathbf{w}_{i,t}}} \right) + {c_2}\left( {\mathbf{w}_t^{\overline{ g} } - {\mathbf{w}_{i,t}}} \right) & \nonumber \\
    && &- \alpha \nabla F\left( {{\mathbf{w}_{i,t}}, D_g} \right), & \label{eq:DSL-OTA_update}  \\
    &&  {\mathbf{w}_{t + 1}} &= \mathbf{w}_{t} + \frac{1}{{\sum\limits_{i = 1}^C {{s_{i,t + 1}}} }}\sum\limits_{i = 1}^C {{s_{i,t + 1}}\left({\mathbf{w}_{i,t + 1}} - {\mathbf{w}_{i,t}}\right)}. &\label{eq:multi_updates}
\end{flalign}
Leveraging the network intelligence, partial worker participation offers a low power approach to harnessing sufficient data for distributed training with limited communication overhead.


\subsection{Over-the-Air Aggregation model}

Numerous participants incur additive time and frequency costs of information transmission, 
which may lead to the training delay with limited wireless resources. OTA communication model publishes an efficient and secure communication mechanism for uplink transmission. Due to the waveform superposition properties of MAC, selected workers can transmit parameters simultaneously on the same channel~\cite{csahin2023survey}.  This multiple access method provides a naturally summed signal for the receiver, which keeps the PS away from sensitive individual information. 
OTA aggregation not only performs well with the malicious interference via synchronous transmission,  but paves the way to feasible multiple access solutions for resource conservation~\cite{fan2021joint}. 

In this case, OTA aggregation is defined based on the individual model parameter vector $\mathbf{w}_{i,t}=\{w_{i,t}^n\}, i=1, \ldots,C$ to update the global model parameter $\mathbf{w}_{t}=\{w_t^n\}$. A time-varying channel model $h_t$ is designed for the transmission of uplink signals in wireless communications. With the launching signals set $\{\mathbf{w}_{i,t}\}$, the received signal $\mathbf{y}_t $ is depicted as
\begin{flalign}
    &&   \mathbf{y}_t &= h_t\left( {{\mathbf{w}_{i,t}};{\mathbf{z}_t}} \right) & \nonumber \\
    &&     &= \sum\limits_{{s_{i,t}} \in {S_t}} {{\mathbf{p}_{i,t}} \odot {\mathbf{w}_{i,t}} \odot {\mathbf{h}_{i,t}}}  + {\mathbf{z}_t}  & \nonumber \\
    &&     &= \sum\limits_{{s_{i,t}} \in {S_t}} {{\mathbf{b}_{t}} \odot {s_{i,t}} \odot {\mathbf{w}_{i,t}}}  + {\mathbf{z}_t}, & \label{eq:channel_coding} 
\end{flalign}
where $h_t\left(\cdot\right)$ represent the channel model of wireless communications. $\mathbf{p}_{i,t}=\left[ {p_{i,t}^1, \ldots ,p_{i,t}^N} \right],p_{i,t}^n=\frac{s_{i,t}b_t^n}{h_{i,t}^n}$ reveals the transmitter control scalar of $i$-th device. The power scaling factor $\mathbf{b}_{t} \in \mathbb{C}$ represents the intrinsic emission capacity of devices. The channel gain of user-$i$ at time-$t$ $\mathbf{h}_{i,t}$ represents the influence of MAC on signal transmission. $\mathbf{z}_t \sim  \mathcal{C}\mathcal{N} \left( {0,\sigma {I^2}} \right)$ is an additive withe Gaussian noise (AWGN). 

The PS approximates the summed $\mathbf{w}_t$ from  $\mathbf{y}_t$ as
\begin{flalign}
    &&  {\mathbf{w}_t} &= {\left( {S_t \odot {\mathbf{b}_t} } \right)^{{ \odot ^{ - 1}}}} \odot {\mathbf{y}_t}   & \nonumber  \\
    &&  &= {\left( {S_t} \right)^{{ \odot ^{ - 1}}}}S_t \odot {\mathbf{w}_{i,t}} + {\left( S_t \odot {\mathbf{b}_t}  \right)^{{ \odot ^{ - 1}}}} \odot {\mathbf{z}_t}. & \label{eq:decoder}
\end{flalign}
Based on the averaging aggregation method~\cite{mcmahan2016federated}, we introduce a balanced cooperation to ensure participants share similar significance in parameter aggregation by the scaling vector ${\textstyle{\textstyle{( {\sum\limits_{{s_{i,t}} \in {S_t}} {{s_{i,t}} \odot {b_t}} })^{{ \odot ^{ - 1}}}}}}$, where $\left(\cdot \right)^{{ \odot ^{ - 1}}}$ represents the inverse Hadamard-product operation. Hence, the PS get $\mathbf{w}_t$ without any prior knowledge about $\mathbf{w}_{i,t}$. OTA satisfies the security and privacy concerns of DSL, and performs as an access multiplexing communication model for the multi-worker selection based DSL. 

\subsection{Joint Optimization of Worker Selection and Over-the-Air}
By collaborative optimizing the worker selection and OTA limitations, DSL-OTA figures out the efficient and comprehensive devices collaboration in wireless systems. The communication constraint is characterized by the requirement that the transmitter must provide sufficient power for a discernible uplink signal limited by the device's maximum energy storage. 
\begin{flalign}
    &&   {\left| {{p_{i,t}w_{i,t}}} \right|^2} &= {\left| {\frac{{{s_{i,t}}{b_t}}}{{{h_{i,t}}}}{w_{i,t}}} \right|^2} \le {P_i^{\max}},  & \label{eq:OTA_constrain} 
\end{flalign}
where $\{P_i^{\max}\}$ represents the maximum transmit power of devices.

Based on the submultiplicative property of norms $\left| {{\bf{XY}}} \right| \le \left| {\bf{X}} \right| \cdot \left| {\bf{Y}} \right|$, we have the sufficient condition of \eqref{eq:OTA_constrain}.

 \begin{flalign}
    &&   {\left| {\frac{{{s_{i,t}}{b_t}}}{{{h_{i,t}}}}} \right|^2} {\left|w_{i,t}\right|^2} &\le {P_i^{\max}}.  & \label{eq:imp_OTA_constrain} 
\end{flalign}
Since the target model $M$ is fixed, individual parameter transmission process share the same scale of signal ${\left|w_{i,t}\right|^2}$. \eqref{eq:imp_OTA_constrain} is known as the wireless deployment constraint of DSL-OTA for recognizable information interaction.

Consider wireless communication of the distributed learning, joint optimization  of DSL-OTA is formulated as

\begin{flalign}
    &&  J(S_t) =& \max{\sum\limits_{{S_t}} {{s_{i,t}}}}   & \label{eq:DSL-OTA_optimization} \\
    &&  s.t.& {\left| {\frac{{{s_{i,t}}{b_t}}}{{{h_{i,t}}}}} \right|^2} {\left|w_{i,t}\right|^2} \le {P_i^{\max}} & \nonumber \\
    &&  &  \theta_{i,t}s_{i,t} \le \overline \theta  _{t-1} ,i = 1, \ldots ,C & \nonumber \\
    && {s_{i,t}} & \in \left\{ {0,1} \right\}. & \nonumber 
\end{flalign}

We figure out the optimal solution for the non-convex problem \eqref{eq:DSL-OTA_optimization} by the augmented Lagrangian method (ALM), and get reliable solutions with stable searching process. The joint optimization of multi-worker selection and OTA aggregation achieves proper participants selection based on dataset, validation accuracy and energy constraint for limited wireless resources and low power communications.

To fully demonstrate the learning and communication model, the implementation of our DSL-OTA is described in Algorithm~\ref{alg:DSL-OTA}. 

\begin{algorithm}[!htb]
	\caption{DSL-OTA}
	\label{alg:DSL-OTA}
	\begin{algorithmic}[1]
\renewcommand{\algorithmicrequire}{\textbf{Initialization:}}
		\REQUIRE ~~\\
		Init $\mathbf{w}, \mathbf{w}_{i,t}^l,\mathbf{w}_{i,t}^{\overline{g}}$, $t,i$. Dataset $\left\{ {{D_i}} \right\}, D_g$ \\
        
        Calculate the data consistency $\{\eta_i\}$; \\
    \FOR {each communication round $t=1:T$}\vspace{0.08in}

        \STATE \!\!\!\!\textbf{at each local worker:$i \in C$}
            \STATE model the wireless communications $\mathbf{z}_t,\mathbf{b}_t, \mathbf{h}_t $, and determine $P_{i}^{max}$ ;
            \STATE$\mathbf{w}_{i,t}^l \leftarrow \mathop {\arg \min }\limits_{w\in\{w_{i,t-1},w_{i,t}\}} \left\{ {F_{i,t-1},F_{i,t}} \right\} $ ;
            \STATE\hspace{0.1in} update $w_{i,t + 1}$ via \eqref{eq:DSL-OTA_update}, calculate $F_{i,t + 1},F_{i,t + 1}^g$ via \eqref{eq:RMSE} for training and evaluation; 
            \STATE\hspace{0.1in}  formulate the uplink channel of  $i$-th worker and the PS by \eqref{eq:channel_coding}, derive the power constraint of the channel via \eqref{eq:imp_OTA_constrain};
            \STATE\hspace{0.1in} search for the set of better workers $S_{t + 1}$ via the multi-worker selection optimization function \eqref{eq:DSL-OTA_optimization}; 
            \STATE\hspace{0.1in} upload $\left\{ {{\mathbf{w}_{i,t + 1}},{F_{i,t + 1}}} \right\},i \in {S_{{{t + 1}}}}$ to update $\mathbf{w}_{t+1}$ via \eqref{eq:multi_updates};
        \vspace{0.1in}

        \STATE \!\!\!\!\textbf{at the PS:}
         \STATE\hspace{0.1in} decode the summed parameters ${\mathbf{w}_{{{t + 1}}}} $ via \eqref{eq:decoder}; 
        \STATE $\mathbf{w}_{i,t}^{\overline{g}} \leftarrow   \mathop {\arg \min }\limits_{\mathbf{w}\in\{\mathbf{w}_{t-1},\mathbf{w}_{t}\}} \left\{{F_{t-1},F_{t}} \right\} $; 
        \STATE\hspace{0.1in}{broadcast $\mathbf{w}_{i,t}^{\overline{g}}$ to all workers.}
    \ENDFOR
	\end{algorithmic}
\end{algorithm}


\section{Theoretical Analyses} 
This section analyzes convergence behaviors and communication efficiency of DSL-OTA, indicating an 
affordable convergence rate and scalable 
saving on transmission resources. Since the page limit, this paper briefly introduce the convergence performance and detailed the communication efficiency of DSL-OTA in wireless networks, we will provide the entire analysis in the journal version of this paper. 

DSL-OTA achieves a comparable convergence rate $ \mathcal{O}\left(1 \right)$ as other MAC based communication models~\cite{fan2021joint, oh2024communication}. The implementation of ALM algorithm does not incorporate additive training workload as the convergence rate $\mathcal{O}\left(\frac{1}{k}\right)$ serves as the higher order infinitesimal of DSL-OTA algorithm for a large fitting round $k$.  
The communication overhead of distributed learning systems can be defined as the requirements of time and frequency resources in wireless communication uplink~\cite{shahid2021communication}. Assuming the standard FedAvg achieves wireless communications via the orthogonal frequency domain multiple access (OFDMA), the Shannon capacity based recognizable signal rate is~\cite{
elbir2021federated} 

\begin{align}\label{eq_1:shannon}
r_{k,t}= B _{k,t} {\log _2}{\left(1+\frac{h _{k,t} p _{k,t}}{{P_z B _{k,t}}}\right)}, \; k=1, \ldots, C
\end{align}
where for the uploading channel between worker $k$ and the PS, $B_{k,t}$ represents the bandwidth of given channel at $t$ communication round,  $P_z$ is the power spectral density of noise $\mathbf{z}_t$.To upload the fixed size of uplink signal $\|\mathbf{w}\|$, the transmission time cost can be defined as

\begin{align}\label{eq_1:time_cost}
T_{k,t}\ge \frac{\|\mathbf{w}\|}{r_{k,t}}.
\end{align}

 Hence the communication cost in time $T_t$ and bandwidth $B_t$ of standard FL in $t$-th communication round is  
 \begin{align}\label{eq_1:time_frequency_cost_FL}
T_t &= {\frac{{\left\| {\bf{w}} \right\|}}{{{ \underline{B}_{t}}{{\log }_2}\left( {1 + \frac{{{h_{k,t}}{p_{k,t}}}}{{{P_z}{\underline{B}_{t}}}}} \right)}}} \nonumber\\
B_t &= \sum\limits_{k = 1}^C {B_{k,t}},
\end{align}
where $\underline{B}_{t}=\min\limits_{i \in C} \{B_{i,t}\}$ represents the minimal frequency overhead of selected workers. $T_t$ leads to the time cost of the parameter uploads process with frequency multiplexing.

DSL-OTA allows participants transmit their parameters simultaneously via OTA aggregation. The communication overhead is 
 \begin{align}\label{eq_1:time_frequency_cost_DSL-OTA}
T_t &=  {\frac{{\left\| {\bf{w}} \right\|}}{{{ \underline{B}_{t}}{{\log }_2}\left( {1 + \frac{{{h_{k,t}}{p_{k,t}}}}{{{P_z}{{ \underline{B}_{t}}}}}} \right)}}} \nonumber\\
B_t &= \max\limits_{i \in S_t} \{B_{i,t}\}. 
\end{align}
where $\underline{B}_{t}=\min\limits_{i \in S_t} \{B_{i,t}\}$. For the equal bandwidth carrier division, $B_{k,t}=B_t$. 

DSL-OTA shares the similar time cost as FedAvg, and achieves $\frac{1}{C}$ of the frequency cost than FedAvg while training the same model. Compared to DSL, DSL-OTA accomplishes cooperative learning of multiple participants as the similar communication overhead as the single worker framework.

\section{Simulation Results}
In this section, simulation results are evaluated to highlight the improvement of DSL-OTA in wireless communication networks. Compared with the benchmark DSL~\cite{fan2023cb} and FedAvg~\cite{mcmahan2016federated}, DSL-OTA achieves faster convergence and higher learning accuracy.

\subsection{Experiments Settings}
We implement DSL-OTA in wireless networks with various workers, each worker contains a individual dataset ${\left| {{D_i}} \right|}=512$ and the synthetic evaluation dataset ${\left| {{D_g}} \right|}=2048$. Systems are designed with different scale of participants $C=10,20,30,40,50$ to verify the communication efficiency of OTA. Local datasets are the time-invariant subset sampling from public datasets CIFAR10 ~\cite{krizhevsky2009learning}. Heterogeneous datasets are generated by the Dirichlet distribution~\cite{hsu2019measuring} with parameter $\alpha=1$, indicating individual data usually consist of different labels. The image classification model trained by DSL-OTA is the ResNet18 ~\cite{casella2024experimenting}. We introduce the SGD optimizer for distributed learning with learning rate $\alpha_{init}=0.01$. The training process involves $40$ communication round, each round includes $10$ epoch. The batch size is $64$. The hyperparameters in PSO is sampled from $c_0 \sim U\left( {0,1} \right)$ and $c_1,c_2\sim \mathcal{N}\left( {0,1} \right)$. The weight hyperparameter $\tau=0.9$ in \eqref{eq:weighted_constrain} represents the influence of data quality and training performance on the worker evaluation. 

To simulate wireless communications, we post the ideal downlink channel assumption for the receivable broadcast information. The uplink channel of $C$ workers is determined by a standard MAC (OTA-based algorithm) or OFDMA (other algorithm) with factors as the power scaling factor $\mathbf{b}_{i,t}\sim \mathcal{N}\left( {{P^{\max }},0.625} \right)$, the channel gain ${\mathbf{h}_i} \sim \mathcal{N}\left( {i,0.625} \right)$ and the AWGN ${\mathbf{z}_t} \sim \mathcal{N}\left( {0,1} \right)$. Distributed devices share the same maximum transmit power ${P^{\max }}=1$. The signal transmission is performed without any delay.

\begin{figure}[!t]
    \centering
    \subfloat[]{\includegraphics[width=1.68in]{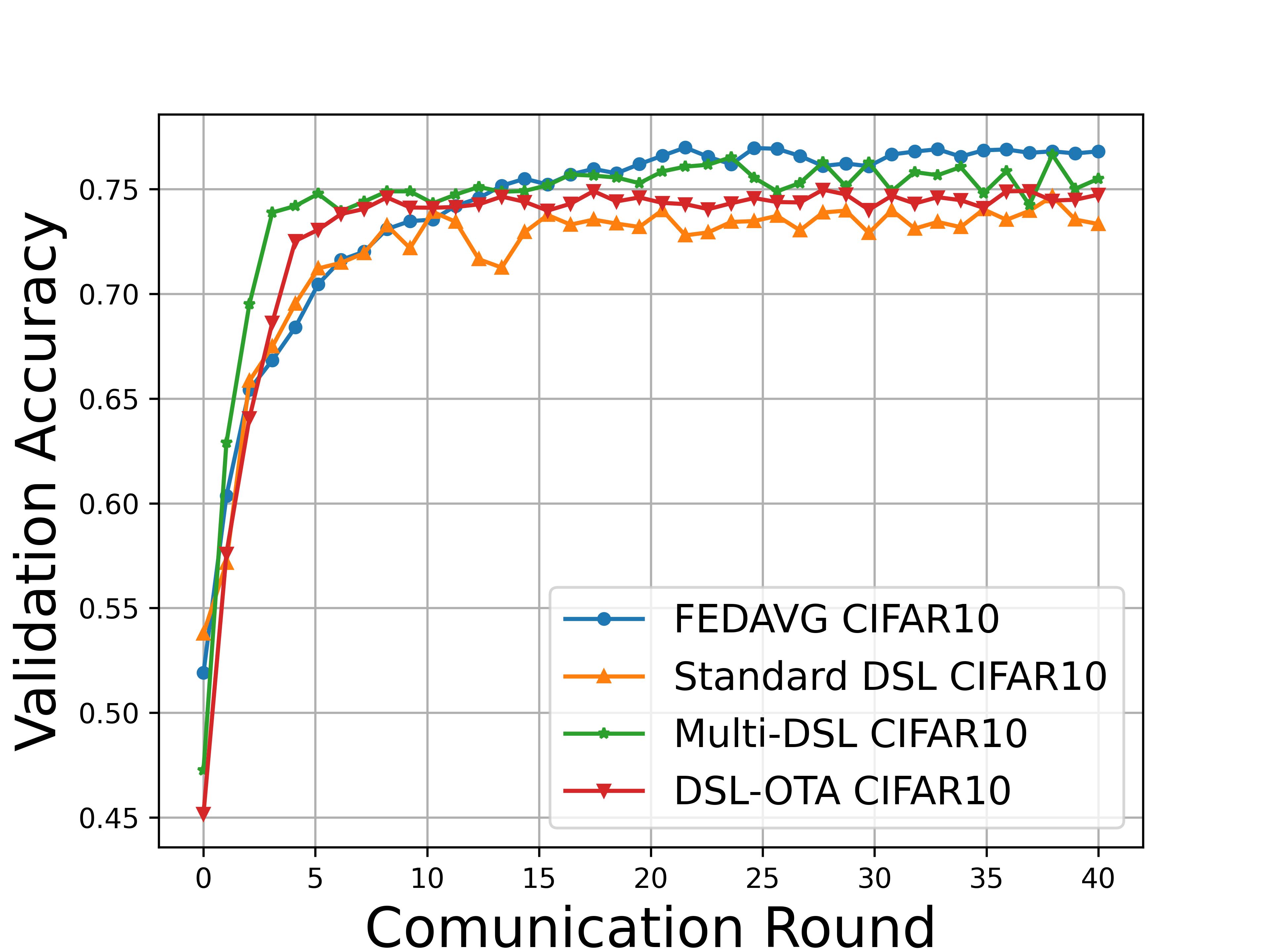}\label{results:iid_methods_compare}}%
    \hfil
    \subfloat[]{\includegraphics[width=1.8in]{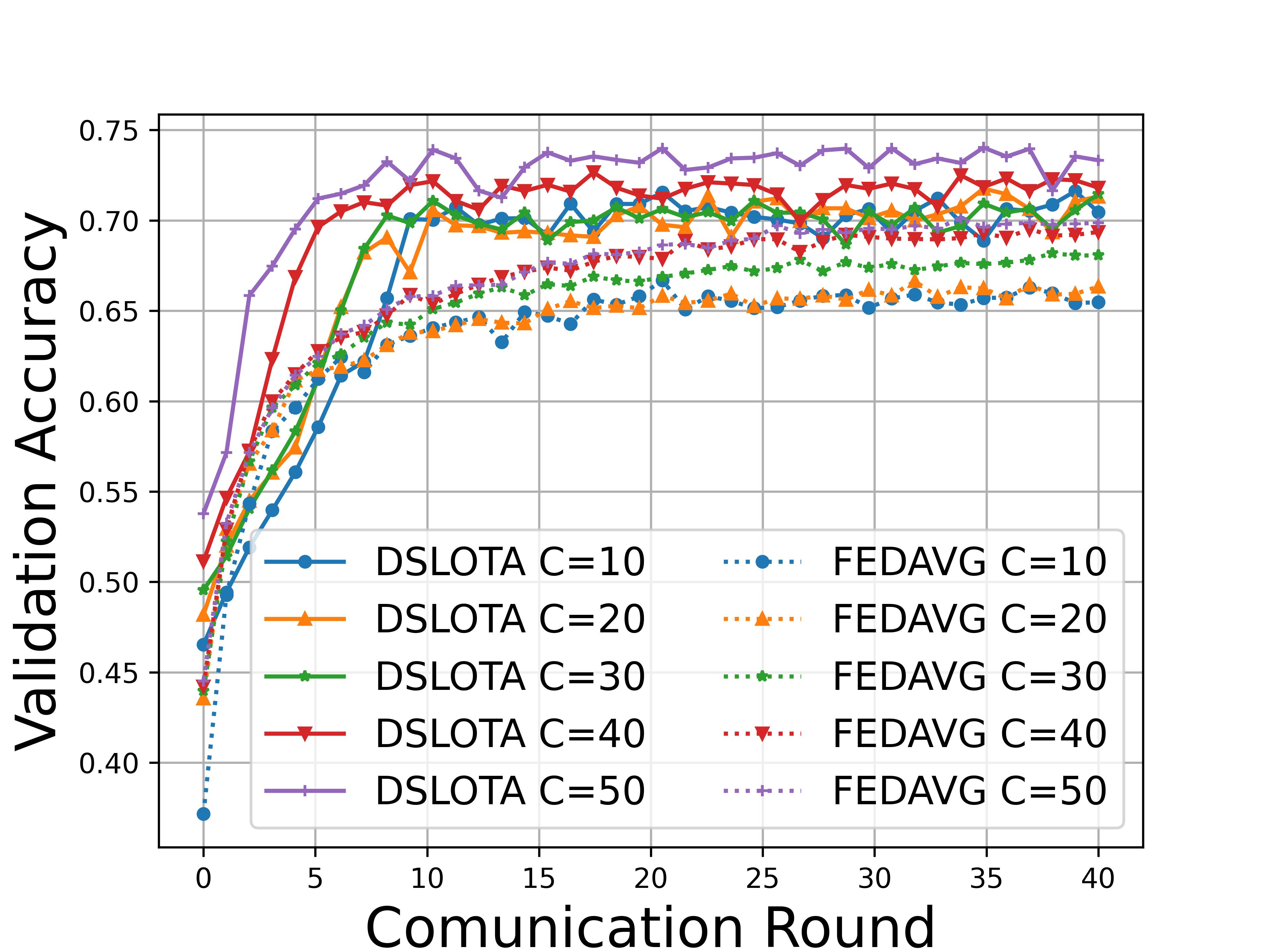}\label{results:iid_scale_compare}}%
    \newline
    \subfloat[]{\includegraphics[width=1.65in]{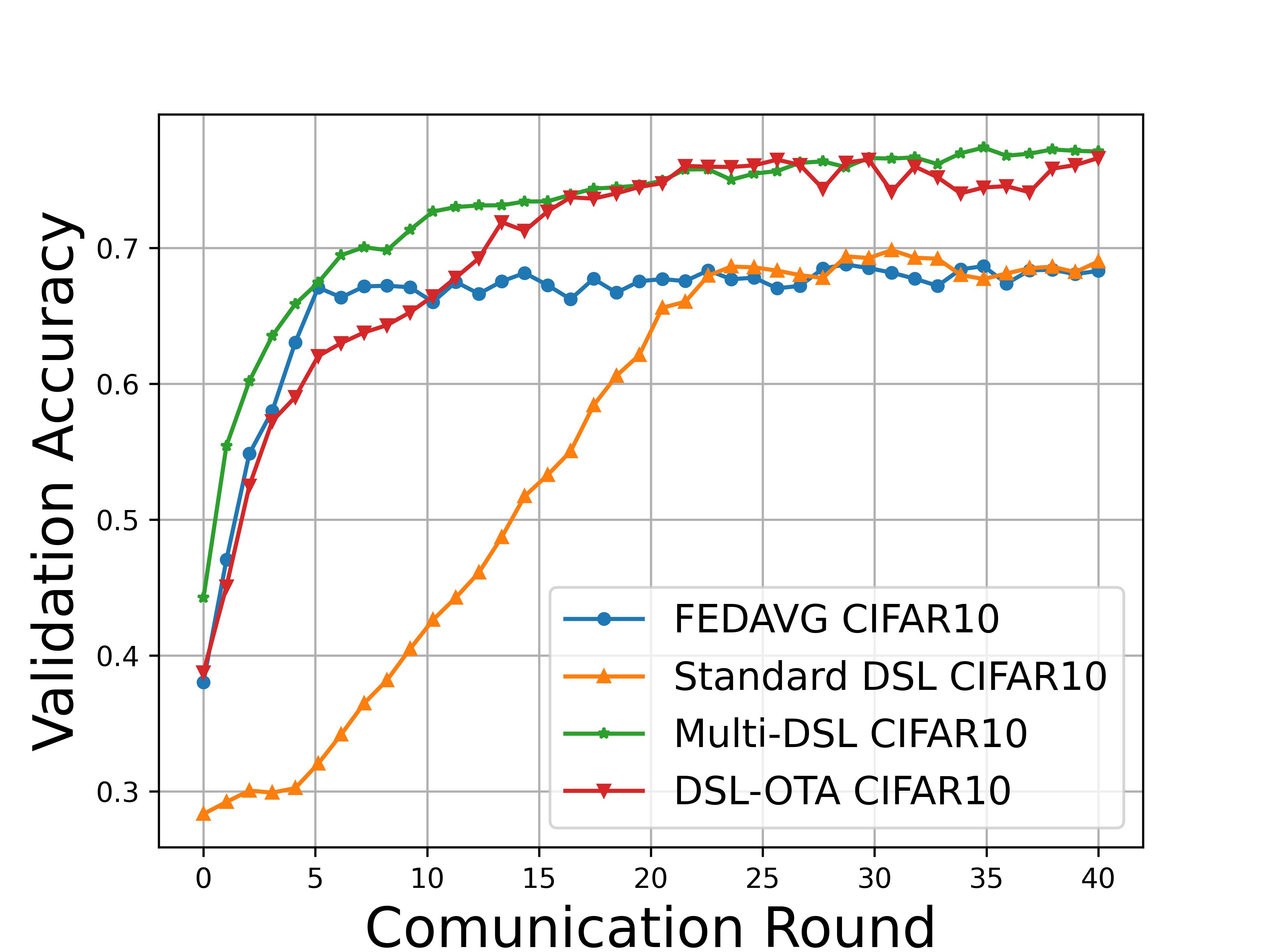}\label{results:noniid_methods_compare}}%
    \hfil
    \subfloat[]{\includegraphics[width=1.78in]{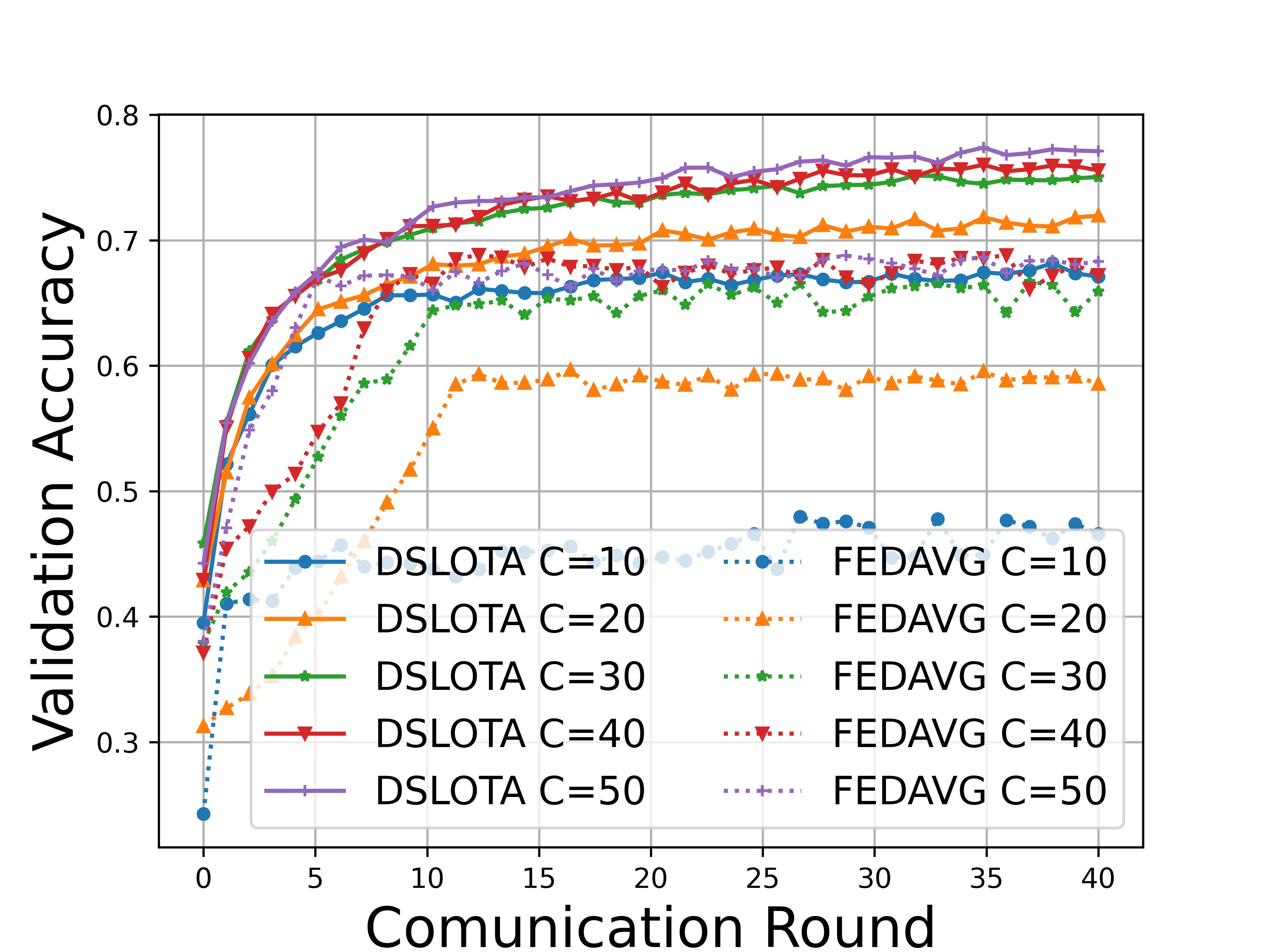} \label{results:noniid_scale_compare}}%

    \caption{Comparative experiments of image classification in wireless systems with data homogeneous and heterogeneous cases.}  
    \label{results}
\end{figure}

\subsection{Performance Comparison}
Comparison studies are designed to verify the superiority of our proposed DSL-OTA beyond FedAvg~\cite{mcmahan2016federated}, standard DSL~\cite{fan2023cb} and DSL-based algorithms with heterogeneous and homogeneous datasets. 
We record the learning accuracy to infer the training effects as Fig. \ref{results:iid_methods_compare} (homogeneous case) and Fig.\ \ref{results}\subref{results:noniid_methods_compare} (heterogeneous case, $\alpha=1$). By introducing the multi-worker selection strategy, the distributed learning progress is improved in the convergence rate and the converged accuracy level, especially in data heterogeneous scenarios Fig.\ \ref{results}\subref{results:noniid_methods_compare}. Despite the involvement of OTA aggregation hinders the converging process at the beginning, DSL-OTA achieves comparable accuracy to the OFDM based wireless networks with less communication costs. 
In Fig. \ref{results:iid_methods_compare} and Fig.\ \ref{results}\subref{results:noniid_methods_compare}, we present how OTA influences convergent behaviors on different scales of realistic wireless communication networks. As the wireless network enlarges, the training process achieves higher accuracy in each round and eventually flats out with different stable states,  since more patterns contributes to global training. However, more participants invoke additional cost. Fig.\ \ref{results}\subref{results:noniid_methods_compare} indicates that heterogeneous networks with fewer participants may miss some significant characteristics for classification. Compared to the OFDM based FedAvg, our DSL-OTA converge faster and achieve outstanding steady accuracy with only one MAC channel occupied.

\section{Conclusion}
In this work, we integrate the OTA aggregation model with the multi-worker selection based DSL technique to enhance 
communication efficiency of wireless networks and to boost collaboration gain among edge devices. To do this, this paper investigates the
, by developing a novel DSL-OTA algorithm. 
Theoretical proof confirms the superiority of DSL-OTA in communication efficiency and convergence guarantee. 
Simulation results verify its capability in desired learning accuracy at faster convergence speed and lower communication costs. 
This work sheds lights on the introduction of multiple access techniques into the mobile edge AI for wireless communications,  leading to  effective collaboration and aggregation schemes under the constraint of limited communication resources. 

\bibliographystyle{IEEEtran}
\bibliography{bib}

\end{document}